# Interaction between stimulated current injection and polariton condensate

Burcu Ozden*[1a], David M. Myers[a], Mark Steger[b], Ken West[c], Loren Pfeiffer[c], David W. Snoke[a]
[a]Department of Physics and Astronomy, University of Pittsburgh, Pittsburgh, PA 15260, USA;
[b]National Renewable Energy Lab, 15013 Denver West Parkway, Golden, CO 80401; [c]Department of Electrical Engineering, Princeton University, Princeton, NJ 08544, USA

## ABSTRACT

In this paper, we see a strong effect of the injected current on the light emission from the polariton condensate in an n-i-n structure, when we monitor the luminescence intensity under applied bias at various pump powers. We present here three thresholds for nonlinear increase of the intensity. We show that small changes of the incoherent injected current lead to stimulated enhancement of the coherent light emission from free carriers. We conclude that the polariton condensate-current system is a highly nonlinear electro-optical system.

**Keywords:** Exciton-polaritons, Bose-Einstein condensation, photoluminescence, nonlinear optics, microcavity, coherent excitation,

## 1. INTRODUCTION

Cavity exciton-polaritons in semiconductor microcavities are quasiparticles resulting from strong coupling between microcavity photons and quantum well (QW) excitons[1]. In the past few decades, there has been an enormous research for the study of optical properties of exciton polariton condensates in semiconductor microcavities. This is due to their desired characteristic of having light mass[2] and manifestation of Bose-Einstein condensates[3,4], superfluity[5], and polariton lasing[6]. However, there has been little effort to study the effects of in-plane transport on polariton condensation, which may be important for optical devices to be used in coherent optical circuits.

Schineder *et al*.[7] and Bhattacharya *et al.*[8] studied electrically pumped polariton lasers studied in *p-i-n* structures. Lagoudakis *et al.*[9] showed exciton-polaritons can interact with carriers when free electrons and holes are introduced in the same in the same QW with the excitons or in a neighboring QW. Brodbeck *et al*.[10] distinguished photon lasing from polariton lasing by monitoring the free carrier reservoir in a GaAs-based quantum well microcavity. It is expected to see some peculiar optical features when exciton-polaritons and free carriers are confined together.[11]

In this paper, we report the results of a study of the effect of injection of free carriers of one polarity into a polariton condensate, in an *n-i-n* structure in which the polariton condensate was confined in a pillar. We probed the luminescence intensity of exciton-polaritons for different pump powers under applied bias between lateral contacts deposited directly on the quantum wells of a microcavity. Concurrently, we investigated the condensate intensity in the traps at the corners and edges of the pillar as a function of optical pump power. Previously we have developed a post-growth etched trapping method[12] which results in a reduced lower polariton energy near the edges of the pillar, independent of the repulsive potential at the location of the pump spot. In these samples, the condensates initially form at the corners of the pillar and then extend across the pillar as the polariton density is increased[12]. The results presented in this report are from the etched pillars similar to in Ref. 12.

We identify three clear thresholds in the photoluminescence signal of the exciton-polaritons as a function of the incoherent pump power. The first threshold occurs when polariton condensation occurs at the optical excitation spot in the middle of the pillar. A second threshold occurs when the condensate appears both in the corners and at the edges of the pillar. The third threshold is new, and has been seen only in these pillars; it occurs when the various trapped condensates in the corners of the pillar begin to spread out of their traps into the regions between the corner traps, and become phase-locked to each other.

---

[1] burcu2@pitt.edu; phone:1 412 624-7861; fax: 1 412 624-9163 http://www.phyast.pitt.edu/~snoke/

The current dependence of the photoluminescence shows a distinct jump at the second threshold, resulting an enhancement in the luminescence intensity of the trapped condensate. There are two physical mechanisms that can lead to this behavior, which we discuss below.

## 2. EXPERIMENTS

The GaAs-based microcavity used in this study was grown by molecular beam epitaxy (MBE) method on a GaAs (001) substrate. The microcavity consists of a $3\lambda/2$ cavity with 32 pairs of AlAs/Al$_{0.2}$Ga$_{0.8}$As $\lambda/4$ layers in the top DBR, and 40 pairs of those layers in the bottom DBR. Inside the cavity, three sets of 4 pairs of 7 nm thick undoped GaAs quantum wells, with AlAs barriers, are placed at the antinodes of the cavity mode, maximizing the coupling with the cavity photons. During the growth process, the thickness of the microcavity is tapered along the radial direction of the wafer leading to a gradient in photon energy towards the edge of the wafer, with about 15% thickness variation from the center to the outside of the wafer.

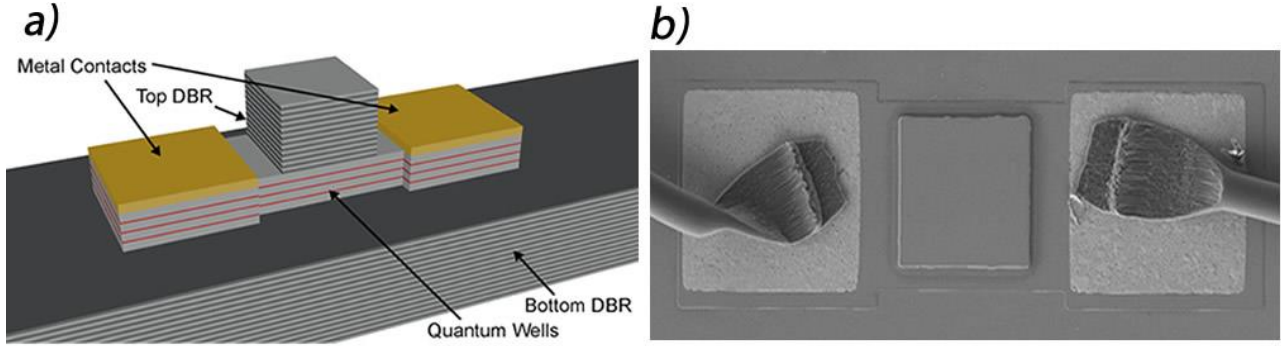

Figure 1. a) The structure of the polariton microcavity with symmetric n-i-n current injection. The distributed Bragg reflector (DBR) mirrors are made of alternating layers of AlAs and Al$_{0.2}$Ga$_{0.8}$As with 40 periods on bottom and 32 periods on top; the QWs are in three sets of four and are pure GaAs/AlAs, and the contacts are both a Ni/AuGe/Ni/Au stack. b) Scanning electron microscope (SEM) image of a typical structure.

For device fabrication, the 3-inch GaAs microcavity wafer was diced into 5 mm x 5 mm individual pieces. A piece from the location on the wafer where the cavity photon was resonant with the quantum well exciton line (1.601 eV) was used to fabricate an array of devices with two n-type contacts on each side of a 100 μm x 100 μm pillar isolated with mesas using standard photolithography techniques. A diagram of the device is shown in Fig. 1a, and a top view taking using a scanning electron microscope (SEM) image is shown in Fig. 1b. The pillars and mesas were dry etched in two stages, using 20:7 BCl$_3$/Cl$_2$ inductively coupled plasma (ICP) reactive ion etch (RIE) at 3.0 mTorr chamber pressure, 600 W ICP power, and 75 W RF bias power. A ~ 2.7 μm photoresist mask was used to etch through the ~ 4 μm thick top DBR and expose the quantum well layers. The mesas were formed by a second etch that was deep enough to remove the quantum wells (~ 400 nm) and ensure electrical isolation of the device. For electrical injection of carriers, square (125 μm x 125 μm) n-type contacts were evaporated on top of the undoped GaAs/AlAs quantum wells on opposite sides of the pillar, made of 5 nm Ni, 120 nm AuGe, 25 nm Ni, and 50 nm Au. In order to ensure diffusion of the contacts into all the layers of quantum wells, the contacts were annealed first at 320° C for 60 seconds and then at 410° C for 30 seconds while at a pressure of 1 bar in a 5% hydrogen and 95% nitrogen gas mixture. Our etching method is similar to other methods[13].

An in-plane electric field was applied by two *n*-type contacts in a nominally undoped AlGaAs square pillar structure, such that current flowed in the plane of the motion of the polaritons, parallel to the plane of the quantum wells. A Keithley 2636B source meter was used to sweep the applied voltage between reverse and forward bias and measure current. The surfaces of the microcavity structure were excited by a continuous-wave, stabilized M Squared Ti: Sapphire pump laser, with an excess photon energy of about 100 meV, giving an incoherent injection of excitons and polaritons. To compare optical pump powers for devices with different detunings (photon fraction of the polaritons), the pump laser wavelength was varied in each case to give optimal absorption at the lowest-energy dip in reflectivity of the stop band above the polariton resonance.

Both the pump power and the bias were varied at constant temperature (~ 4 K). A mechanical chopper with a 2.4% duty cycle at 400 Hz was used to prevent overheating. All reported powers are the peak power of each short pulse. The current-voltage measurements were synchronized with the chopped laser pulses. All current values reported in this experiment are the average of at least 250 measurements at each pump power. Both real-space images and spectrally-resolved images were obtained using a CCD camera on the output of an 0.5 m spectrometer.

A total of nine devices were tested, with lower polariton energy ranging from 1.5938 to 1.5989 eV. These are referred to in text as Device 1, Device 2, etc., in the order of their lower polariton energies.

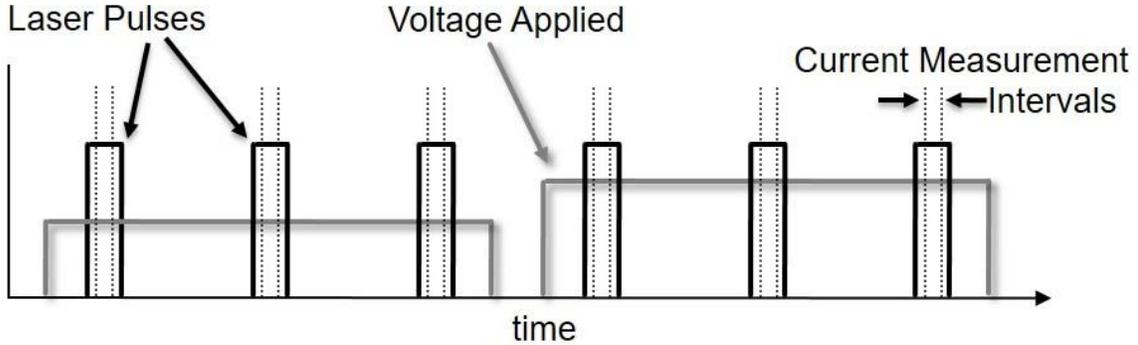

Figure 2. A diagram of the measurement timing used in the experiments for this text. This depicts only three pulses per set voltage, but, it was at least 250 pulses per voltage.

Fig. 2 shows the method of synchronizing the laser pulses and the current pulses used for the current measurements (the pump laser could not be left on continuously, because the sample would become heated.) Quasi-cw laser pulses arrived at a rate of 400 Hz with a width of 60 μs each. A constant DC voltage was applied to the devices at a set voltage for a certain interval, during which an image was taken using a CCD camera with the spectrometer. Simultaneously, current measurements were synchronized to take place during each laser pulse (integration time of 50 μs) with a typical minimum of 250 current measurements for each voltage and pump power. Once one voltage setting was complete, the voltage was increased, a new image was taken, and the current measurement process repeated. When all voltage settings for a given pump power were complete, the pump power was increased and the entire process was repeated at the new pump power.

## 3. RESULTS AND DISCUSSIONS

Fig. 3 shows the condensate intensity integrated over the traps at the corners of the pillar as a function of optical pump power, for one of the devices. There are three thresholds for nonlinear increase of the intensity clearly visible here; in some devices only two of these thresholds are visible when two thresholds are near to each other. This threshold behavior can be understood as follows. Below the lowest threshold, there is a thermal distribution of polaritons concentrated at the pump spot, diffusing throughout the pillar (see Fig. 3b, 10 mW data). The emission at the corners and edges increases linearly with the pump power. The first threshold occurs when condensation occurs at the optical excitation spot in the middle of the pillar. As documented in Refs. [14] and [15], in our long-lifetime structures, as the excitation density is increased, a quasicondensate first appears at the excitation spot. This spot is spectrally shifted to the blue by the potential energy of the exciton cloud on which it sits. The emission of the quasicondensate is spectrally narrow but still has measurable spectral width, and is far more intense than all other regions (see Fig. 3b, 24 mW data). Polaritons from this quasicondensate are not trapped (in fact, they may be viewed as anti-trapped, repelled away by the exciton cloud potential); these polaritons stream freely, ballistically, away from the excitation region. Some of these particles accumulate in the low-energy traps due to partial thermalization, but not enough for condensation in these traps.

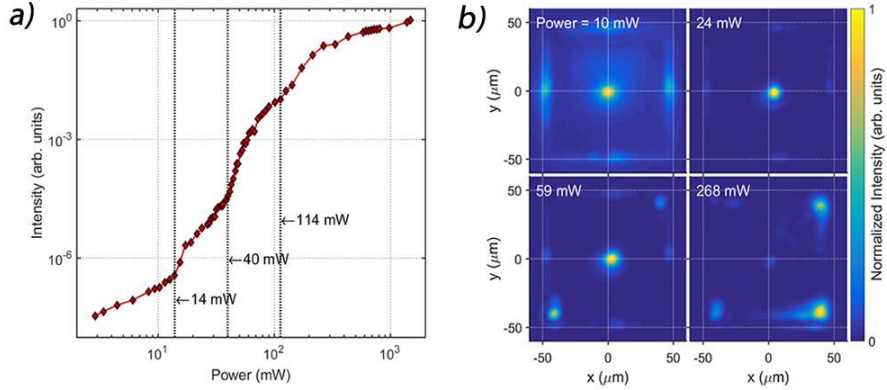

Figure 3. a) Experimental polariton condensate population in the corner traps of the pillar, as a function of incoherent pump power, for Device 3, different from the Device 4 used for the data of Fig. 4 below. b) Full real space polariton luminescence intensity near $k_l = 0$ at different pump powers.

A second threshold occurs when a "true" condensate appears in the traps at the edges of the pillar, namely the corners and edges of the pillar, which are the lowest energy points[12]. This state is characterized by emission which is spectrally very sharp and which is localized to the ground states of the traps (see Fig. 3b, 59 mW data). This behavior has also been seen in other trapping schemes[15,16]. This drop in the photon emission energy, extreme spectral narrowing, and long-distance motion is quite dramatic in the experiments. We tentatively identify this with the onset of superfluid flow into local traps; the superfluid is formed when collisions of polaritons streaming out of the quasicondensate at the laser excitation spot lead to thermalization. At high enough density, the streaming polaritons from the excitation region may have a high enough collision rate with each other to allow scattering into lower energy states, and thereby occupy the trap states at lower energy. The condensates in these traps are well behaved, with well-defined energy.

The third threshold has been seen only in these pillars, and occurs when the various trapped condensates in the corners of the pillar begin to spread out of their traps into the regions between the corner traps (see Fig. 3b, 268 mW data). Eventually, a single monoenergetic condensate, i.e., phase locking of multiple condensates, appears, as reported previously[12]. We report here for the first time the third threshold for nonlinear increase of the total emission intensity, which appears to be associated with the expansion of the trapped condensates out of their traps, leading eventually to contact between the separated condensates. In several of the devices, the lower two thresholds are so close together that they appear as a single threshold, leading to only two clear thresholds, as in Fig. 4c.

Concurrently, we probed the effect of injection entirely of one type of charge carrier into a polariton condensate and showed the enhancement of the luminescence intensity of the condensate. Fig. 4b shows the effect of the current on the light emission from the polariton condensate trapped at the corners of the pillar, under conditions of constant optical pump intensity. Fig. 4d shows the emission intensity from the traps at the corners of the pillar for Device 4 as a function of applied voltage, corresponding to the same data set as the emission intensity versus current data shown in Fig. 4b. The lower polariton energy for Device 4 was 1.5953 eV, corresponding to a detuning of δ = −8.2 meV.

At low optical pump power, when there is no condensate, there is no effect of the current on the light emission. At optical pump power, above the onset for polariton condensation (see Fig. 4c), the light emission from the condensate is dramatically enhanced, by more than an order of magnitude, by the injection of current. At very high optical pump power, when the condensate growth has saturated, there is again no effect of the injected current. This behavior has been seen in several devices with different condensation threshold densities.

At first glance, one would not expect the current to increase the emission intensity, because we have an *n-i-n* structure that injects only free electrons; there are no holes injected to recombine with for light emission. However, this behavior can be explained with two different mechanisms. In one mechanism, there can be some free holes in the pillar due to residual doping or ionization of excitons. When free electrons from an *n*-type region move into the region occupied by the condensate, they can form polaritons by joining a free hole in that region, as illustrated in Fig. 4a. The quantum mechanics of bosons says that every process with a bosonic final state is amplified by a factor $(1 + n_f)$, where $n_f$ is the occupation number of the final state[17]. In polariton condensates of this type, values of the occupation number have been established to at least $n_c = 10$.[18] The exciton-polaritons migrate across the condensate region, and the excitons are ionized into free electrons and holes. Because of the applied voltage across the structure, there is a net motion of the carriers. Free electrons continue out of the system, leading to current in the *n*-type region on the other side, while the free holes travel oppositely to the free electrons, going back to form new polaritons with the injected electrons. Because the polariton condensate does not involve inversion[19], the valence band remains mostly filled and the conduction band remains mostly empty in the range of energies of interest, so that there can easily be motion of free carriers in addition to the motion of polaritons.

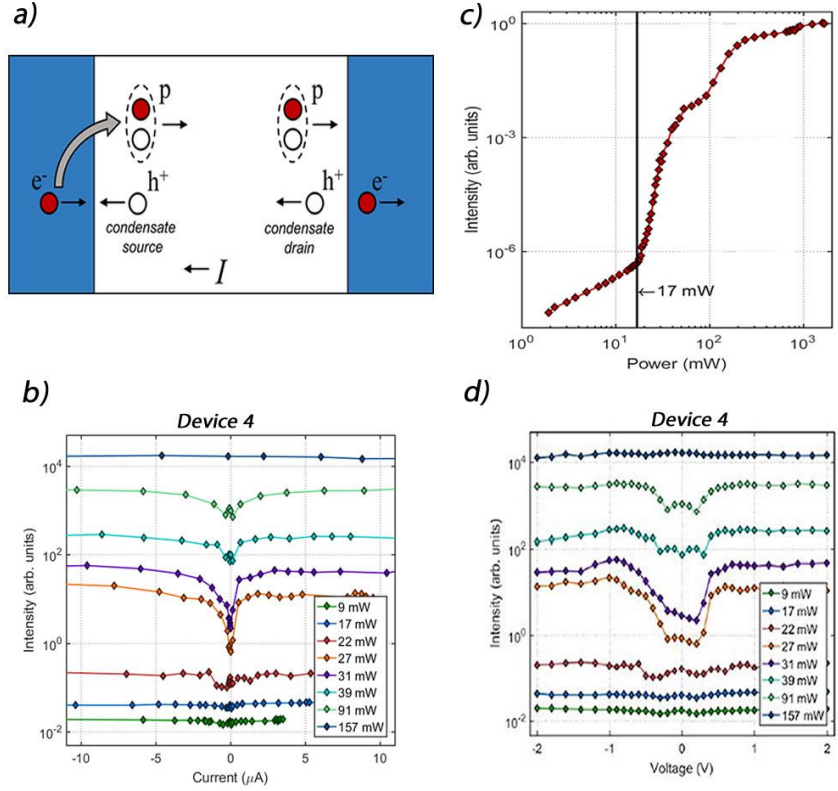

Figure 4. a) Illustration of stimulated creation of polaritons by injection of free electrons, accompanied by spontaneous ionization of excitons, giving a net current from right to left. b) Light emission stimulated by monopolar current injection. The curves are the total condensate emission intensity at the corners of the pillar, as a function of injection current through the pillar at constant optical pump power, for the optical pump powers indicated. c) Condensate emission intensity of the pillar for the data of (b), as a function of optical pump power, with zero voltage across the structure. There are two thresholds visible: the first at low power (here, 17 mW) is for condensation in the corner traps of the pillar. The higher threshold, here at 80 mW, occurs when the separate condensates at the corners expand, to merge into a single extended condensate. d) The emission intensity from the traps at the corners of the pillar as a function of applied voltage.

The effect of the current on the condensate population can be modeled by the following set of rate equations:

$$\frac{\partial n_c}{\partial t} = -\frac{n_c}{\tau_r} + A n_{ex}(1 + n_c) + B n_e n_h (1 + n_c) - \frac{n_c}{\tau_{ion}}$$

$$\frac{\partial n_{ex}}{\partial t} = G - A n_{ex}(1 + n_c) - \frac{n_{ex}}{\tau_{ex}}$$

$$\frac{\partial n_e}{\partial t} = J - B n_e n_h (1 + n_c) + \frac{n_c}{\tau_{ion}} - \frac{n_e}{\tau_{esc}}$$

$$\frac{\partial n_h}{\partial t} = -B n_e n_h (1 + n_c) + \frac{n_c}{\tau_{ion}} - \frac{n_h}{\tau'_{esc}} \tag{1}$$

Here $n_c$ is the polariton population, $n_{ex}$ is the exciton population, $n_e$ is the free electron population, and $n_h$ is the free hole population. The use of an exciton "reservoir" population with stimulated scattering into the polariton population is a standard approach in polariton condensate theory[20], but here we introduce explicitly the role of free carriers as well. The polariton population decays with the radiative time constant $\tau_r$, and can also turn into free electrons and holes with the ionization time constant $\tau_{ion}$, which is temperature dependent. Excitons are generated by the incoherent optical pump with rate $G$, and decay into polaritons with a rate that is stimulated by the final state occupation of the polaritons, with overall multiplier $A$. Free electrons are injected by current $J$, and also can turn into polaritons by picking up a free hole, in a process that is also stimulated by the final state occupation of the polaritons, with overall multiplier $B$. Free electrons can also escape by migration across the surface of the n-i-n barrier with time constant $\tau_{esc}$; holes can also escape, with much lower probability, which is accounted for by making $\tau'_{esc} \gg \tau_{esc}$.

This model can be solved in steady state for the populations as the generation rate $G$ and the current $J$ are increased. When the current $J$ is set to zero, the model reproduces the standard onset of condensation curve at $J = 0$. When the current $J$ is nonzero, the model gives behavior qualitatively the same as that seen in the experiment, as seen in Fig. 5a. We have not explored the parameter variations of this model in detail, because this model is meant to be as simple as possible to show the basic effect. We note simply that the effect of the increase of the light emission with current depends crucially on the stimulated injection term proportional to ($1+n_c$). We emphasize that this effect corresponds to stimulated injection entirely of one of charge carrier, namely free electrons from an n-type region, and not recombination of electrons and holes injected in a p-i-n structure, which could lead to stimulated emission. The injection of free electrons by themselves does not contribute any photons to the system. Rather, the injection of the carriers enhances the capture of pre-existing, uncorrelated holes into correlated polaritons, increasing the rate of recombination leading to light emission.

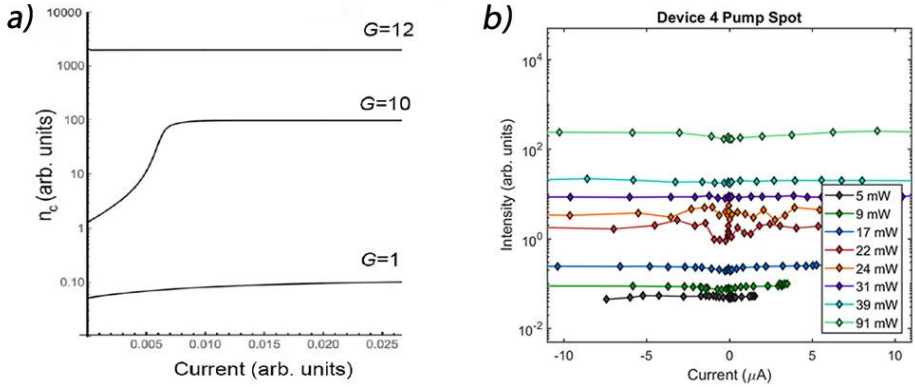

Figure 5. a) Theoretical condensate number $n_c$ as a function of injection current $J$ from the rate equations given in the text, for $A = 10^{-9}$, $B = 0.5A$, $\tau_{ion} = \tau_r = 1000$, $\tau_{ex} = 100\tau_r$, and $\tau_{surf} = \tau_{ex}$, for three optical pump powers $G$ as labeled; $G = 1$ is well below the critical density for condensation, $G = 10$ is at the critical density for condensation, and $G = 12$ is above the critical density b) Condensate emission intensity as a function of current at the pump spot.

A second mechanism could also lead to the effect of the current on the emission intensity of the polaritons. As the current is increased through the structure, the same set of rate equations (1) implies that the density of free carriers can increase. In this case, this polariton-electron scattering can lead to more efficient thermalization of the polaritons. This could move the thresholds discussed above to lower pump power, because the polaritons will have a lower effective temperature at a given density, resulting in the enhancement of emission intensity near a threshold[21]. This effect has been experimentally shown previously for what we call the first threshold, quasicondensation at the pump spot, by other groups[7]. This mechanism is also consistent with the fact that there is no effect on the intensity at very high or low optical pump powers, as shown in Figs 4b and 4c. However, we do not see any effect of the current on the emission intensity at the first threshold, either at the corners of the pillar or at the pump spot (see Fig. 5b). This indicates that this thermalization mechanism is likely negligible in our samples, and the stimulated injection mechanism discussed above is dominant. As discussed above, a key difference in our experiments and those of other groups is the formation of the condensate in traps at the corners and edges of the pillar, more than 50 μm from the pump spot, due to the long-distance motion of the condensate[12], and the jump in PL intensity which we observe occurs only for that trapped condensate, which has high quantum-state occupation number.

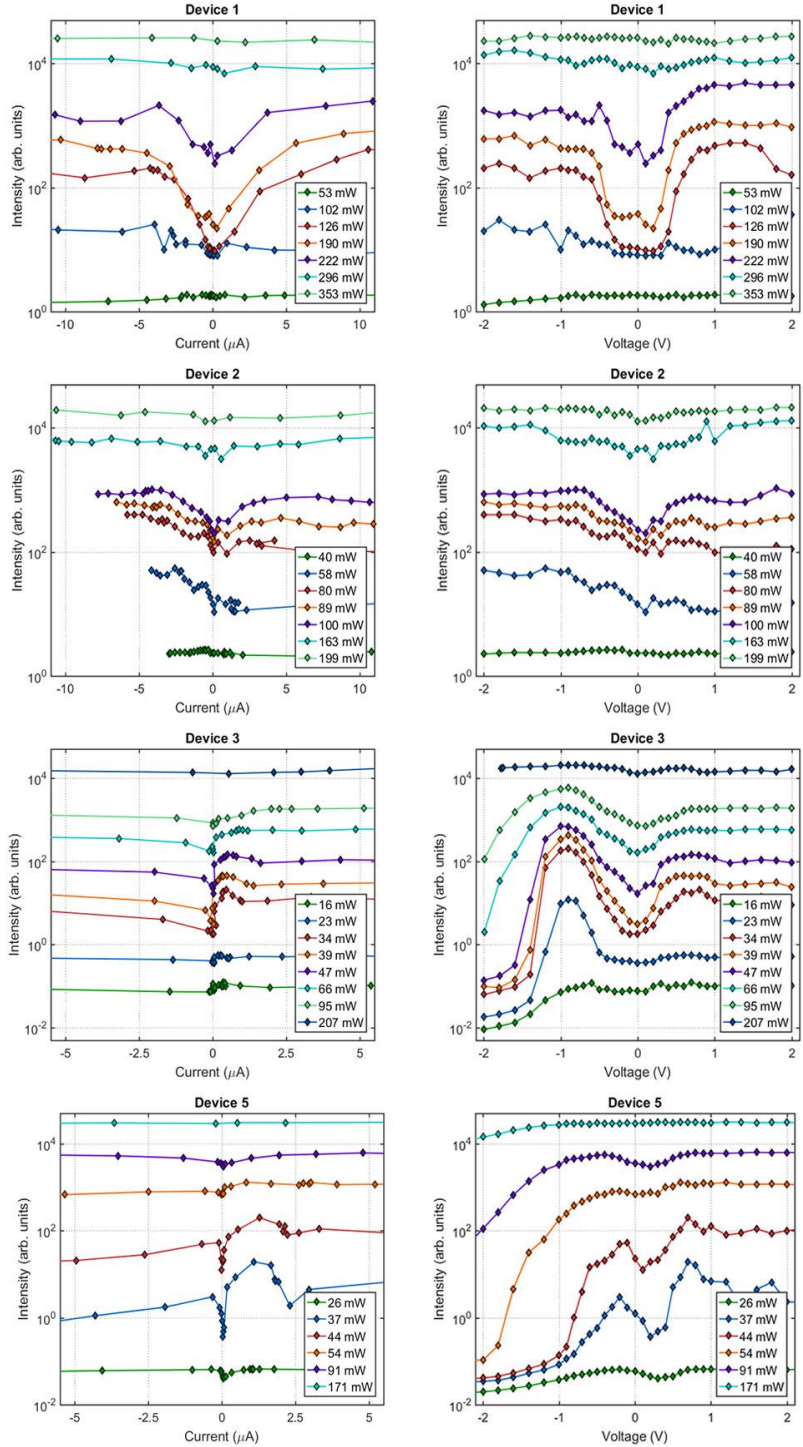

Figure 6. Photoluminescence intensity as a function of current (left column) and voltage (right column) for several devices. The lower polariton energy (LPE) at low density and the condensate threshold powers at 0 V bias are as follows. Device 1: LPE = 1.5938 eV, δ = −10:1 meV, $P_{thres}$ = 150 mW; Device 2: LPE = 1.5942 eV, δ = −9:6 meV, $P_{thres}$ = 50 mW; Device 3: LPE = 1.5948 eV, δ = −8:8 meV, $P_{thres}$ = 40 mW; Device 5: LPE = 1.5961 eV, δ = −7:0 meV, $P_{thres}$ = 29 mW.

A systematic study of the intensity enhancement at different detuning has also been conducted. The detuning $\delta$ is defined as the bare photon energy minus the bare exciton energy; thus, a negative $\delta$ indicates that the polaritons in the device are more photon-like than exciton-like. Fig. 6 shows the emission intensity from the corner traps for several devices, as a function of current and applied voltage. The amount of PL intensity enhancement varies depending on the detuning. As seen in these figures, in some devices at high voltage ($|\Delta V| \geq 0.8$ V), the emission intensity can decrease with increasing voltage (and therefore current). This effect is not restricted to the range of optical pump powers near the condensation threshold. We attribute this effect to increased ionization of excitons due to in-plane electric field. Evidence for this effect has been reported previously[22]. Note that in some cases there is both stimulated injection at low voltage and suppression of the emission due to field-enhanced ionization, at high voltage. We attribute the asymmetries in the characteristics at high voltages to doping differences and inhomogeneities in the *n*-type contacts. For the data of Fig. 6, the number of current measurements per voltage was at least 250 for all devices except Device 5, for which it was at least 150.

## 4. CONCLUSIONS

In conclusion, we have seen that the system of a polariton condensate coupled to in-plane current is a highly nonlinear electro-optical device, in which small changes of the injected current have a dramatic effect on the coherent light emission. Although the injected current is incoherent, it is carried in part of its path by coherent electron-hole pairs involved in the polariton condensate. The evidence supports the conclusion that coherence of the condensate leads to stimulated enhancement of exciton polariton formation from free carriers, although there may also be an effect of increased thermalization due to increase of the free carrier density.

## ACKNOWLEDGEMENTS

This work has been supported by the Army Research Office Project W911NF-15-1-0466. We thank Allan Macdonald and Ming Xie for conversations on the interaction of electrical transport and optical condensates.